\documentstyle[prl,aps]{revtex}

\def\al{{\alpha}}
\def\be{{\beta}}

\def\gam{{\gamma}}

\def\Lam{{\Lambda}}
\def\eps{{\epsilon}}

\def\sig{{\sigma}}
\def\Sig{{\Sigma}}
\def\om{{\omega}}
\def\w{{\omega}}

\def\grad{{\nabla}}

\def\bL{{\mathbf{\Lambda}}}

\def\bT{{\bf T}}

\def\cH{{\mathcal{H}}}
\def\cL{{\mathcal{L}}}

\def\cP{{\mathcal{P}}}

\def\d{{\partial}}

\def\Boxh{\hat{\mbox{\kern-.0em\lower.3ex\hbox{$\Box$}}}}

\def\ddet4g{{\mathstrut^4\! g}}

\def\Lie{{\pounds}}

\def\Kuchar{{Kucha\v r}}

\def\Ham{{\cal H}}

\newcommand\beq{\begin{equation}}
\newcommand\eeq{\end{equation}}
\newcommand\bea{\begin{eqnarray}}
\newcommand\eea{\end{eqnarray}}
\newcommand\beqa{\begin{eqnarray}}
\newcommand\eeqa{\end{eqnarray}}

\begin{document}

\draft
\twocolumn[\hsize\textwidth\columnwidth\hsize\csname
@twocolumnfalse\endcsname

\title{Generalized Einstein theory
with fundamental cosmological stress tensor}

\author{Arlen Anderson}
\address{
Dept. Physics and Astronomy,
Univ. North Carolina,
Chapel Hill NC 27599-3255}
\date{Feb. 10, 1999}

\maketitle
\vspace{-3.5cm}
\hfill IFP-UNC-529

\hfill gr-qc/9902027
\vspace{2.5 cm}

\begin{abstract}
\widetext
Careful analysis of parametrized variational principles in mechanics and 
field theory leads to a generalization of Einstein theory that includes a 
cosmological stress tensor.  This generalization also follows by restricting 
variations of the metric in the Hilbert action to spacetime diffeomorphisms.  
The equation of motion for the generalized theory is the twice-contracted 
Bianchi identity while the field equations constitute the stress tensor of 
the theory. Gravity is interpreted as a cosmological fluid.  
\end{abstract}
\narrowtext
\vskip2pc]

The cosmological constant is popularly regarded as a ``fudge factor'' 
and as Einstein's ``biggest blunder.'' Careful reexamination of 
variational principles for para\-metrized theories, prompted by recent 
results on a modified action principle for general relativity\cite{AnY98}, 
leads to a surprising conclusion that challenges this perception:
the parametrized variational principle for general relativity, and
hence Einstein theory itself, should be
generalized to include not simply a fundamental cosmological constant but 
a cosmological stress tensor.  In particular, the Hamiltonian and 
momentum densities of the gravitational field are not constrained to 
vanish, even in vacuum.  This might appear to be a radical conclusion,
but it is in fact a straightforward consequence of a compelling
approach to parametrized particle mechanics and field theory that
originates in work of \Kuchar\cite{Kuc76,Kuc81}.  While motivated by entirely 
different considerations, some related conclusions have been 
reached previously by efforts to incorporate the cosmological constant into
general relativity by treating it  as an integration
constant\cite{integ,unimod}.

There has been virtually unanimous
agreement following the pioneering work of Dirac \cite{Dir} and
of Arnowitt, Deser and Misner \cite{ADM62} (ADM) that general relativity is
an ``already parametrized'' theory:  the Hamiltonian and momentum
densities in the action are multiplied by Lagrange multipliers, the 
lapse and shift, that when varied produce the Hamiltonian and momentum 
constraint equations.  The crucial assumption that is altered here is the 
identification of the lapse and shift as Lagrange multipliers.  In 
the context of particle mechanics and field theory, I show instead that 
the lapse and shift are more naturally identified as parameters, 
arbitrarily specified, that set coordinates in a fixed
background metric by parametrizing the basis one-forms. 
A true Lagrange multiplier (absent from the conventional formulation of 
general relativity) is needed to impose the definition of these 
parameters.  This Lagrange multiplier is identified with the energy 
in particle mechanics and with the stress tensor in field theory.  
The parametrized action for Einstein theory, being a field theory, thus 
involves a (divergence-free) stress tensor $\bL_{\mu\nu}$, even in the
absence of matter. Variation of the coordinate-fixing parameters
gives 
\beq
\label{Ein1}
G_{\mu\nu} = {1\over 2} \Lam_{\mu\nu}.
\eeq   

An equivalent way to reach this generalization of Einstein
theory is to argue that the
four-metric should not be freely varied in the Hilbert action but
restricted to variations corresponding to (active) spacetime diffeomorphisms.
This is sensible because free variation can take the metric out
of the allowed configuration space, for example, by changing the
signature, while spacetime diffeomorphisms reach all locally accessible
configurations.  The equations of motion that follow from this variation
are well known to be the twice-contracted Bianchi identities. 
A divergence-free stress tensor can be subtracted from these identities
without affecting their validity, and this gives
\beq
\label{bi1}
\grad^\mu(G_{\mu\nu} - {1\over 2} \Lambda_{\mu\nu})=0,
\eeq
where $\grad^\mu$ is the spacetime covariant derivative. 

{}From this perspective, general relativity can be interpreted as 
a theory of a cosmological fluid with stress tensor (\ref{Ein1})
and equations of motion (\ref{bi1}). 
These equations are not a hyperbolic system as they stand. The spatial stresses must be determined in order for the
equations of motion to be definite.  Simply specifying the stresses
everywhere on spacetime presumes too much prior knowledge.
Fortunately, the example of fluid mechanics suggests a resolution
of the difficulty.  The fluid equations of motion are also of
the form $\grad^\mu T_{\mu\nu}=0$. The spatial stresses are determined
by imposing an equation of state in the rest frame of the fluid
to relate pressure and density.  Similarly imposing an equation of
state for the cosmological fluid of gravity completes this
generalization of Einstein's theory. 

Just as in the case of fluids, a preferred
background---the rest frame of the fluid---is implicit to make the
split between spatial and temporal components.  In the
cosmological context, it is natural to conjecture that this
preferred frame is the rest frame of the cosmic microwave background
radiation.  It should be emphasized that the existence of a preferred
frame in no way detracts from the spacetime covariance of the theory.
Indeed, it is spacetime covariance that leads to the recognition
that the Lagrange multiplier must be a two-index tensor.  The
equations (\ref{Ein1}) are first integrals of (\ref{bi1}) in a 
suitable sense and justify the identification of the temporal components
of the Einstein tensor as energy and momenta.

The theoretical possibility of a nonzero cosmological stress tensor
has many applications.  To name two, it may serve as an exotic
source of energy density in the universe, and it might 
resolve the problem of time in quantum gravity. 
Recent experimental evidence from supernovae\cite{supernovae} support the 
conclusion that the energy density of the universe does not arise solely 
from conventional matter and radiation.  To account for this missing
mass, cosmologists have been  
considering alternative sources of energy density, such as 
a nonzero cosmological constant \cite{cosmo} or exotic 
matter \cite{quint}.   A 
cosmological stress tensor can fill this role without the need for
new types of matter (which could only be observed 
indirectly in any event) while allowing for flexibility in
the equation of state. 

One aspect of the problem of time in quantum gravity is that a 
vanishing Hamiltonian density precludes the existence
of an external time variable.  In the Dirac approach, the Hamiltonian
constraint is applied as an operator constraint on states.
For a self-adjoint Hamiltonian, the ostensible Heisenberg
equation of motion leads to the ``frozen formalism'' in which
observables do not evolve in an external time.  In the unimodular approach 
to quantum gravity\cite{unimod}, the Hamiltonian density does 
not vanish but is proportional to the cosmological constant $\Lam$,
which is interpreted as an energy eigenvalue.  An external
cosmological time ({\it cf.} also \cite{cosmo_time}), conjugate 
to $\Lam$, is introduced as an external time, and 
a Schr\"odinger equation for the wavefunction of the universe results.
Similar results follow from the assertion that the ADM action supplemented
by the cosmological constant is a Jacobi action \cite{Jacobi}.
(Evidence regarding this conjecture is given below.)
These approaches foundered
when \Kuchar\ pointed out\cite{Kuc_no} that the cosmological constant does not
have enough degrees of freedom to be conjugate
to the many-fingered time of a field theory.  The cosmological stress tensor 
here overcomes this objection,
and the basic argument of \cite{unimod} may be carried through\cite{And99}. 

We begin with a discussion of the parametrized particle. 
Consider the Hamiltonian form of the action (Hamilton's principal
function \cite{Gol}) for a particle with time-independent Hamiltonian 
in 0+1 dimensions
\beq
\label{action}
S=\int_{\tau'}^{\tau''} [p\d_\tau q - H(p,q)]d\tau.
\eeq
The metric here is $ds^2 =-d\tau^2$.
This metric can be parametrized by introducing the coordinate time $t$ 
through a parametrization of the basis one-form $d\tau$, $ d\tau = N(t) dt$.
Changing variables from $\tau$ to $t$ in the action (\ref{action}) gives 
\beqa
\label{action_t1}
S&=& \int_{t'}^{t''} [p\d_t q - H \d_t \tau] dt \\
\label{action_param}
&=& \int_{t'}^{t''} [p\d_t q - N H] dt.
\eeqa
Varying $N$ in (\ref{action_param}) leads to the absurd conclusion $H=0$.
On the other hand, were $\tau$ varied in (\ref{action_t1}), 
a correct result, $dH/dt=0$, would be found.  This 
realizes Noether's theorem for time reparametrization invariance 
({\it cf.} \cite{Lan}).  

The error made by varying $N$ in (\ref{action_param}) is that $N$ is not free 
but restricted to equal $d\tau/dt$ from parametrization of the metric.  
The relation $N=d\tau/dt$ must be enforced in
the variational principle by imposing it with a Lagrange multiplier
$p^\tau$.   This gives
\beq
\label{action_ext}
S = \int_{t'}^{t''} [p\d_t q  -N H - p^\tau (\d_t \tau -N)] dt 
\eeq
This in turn can be read as an action on an extended phase space of
$(q,p)$ and $(\tau,-p^\tau)$, with a constraint $H-p^\tau=0$ serving
to determine the new degree of freedom $p^\tau$ in the phase space.
In the language of \Kuchar, $\tau$ corresponds to an ``embedding'' variable
that describes how a ``slice'' is embedded in spacetime as a function of
the coordinate $t$.  This derivation differs from the standard argument 
({\it e.g.} \cite{Kuc81}) in which one replaces $H$ in
(\ref{action_t1}) by a momentum $-p^\tau$ and 
then imposes the relation $H+p^\tau=0$ with a Lagrange multiplier $N$.   

Further understanding follows upon exploiting the observation that $\tau$ is
a cyclic variable that can be eliminated from the action by Routh's
procedure \cite{Lan}.  Add the integral of the total 
derivative $\d_t(p^\tau  \tau)$ to $S$ to obtain 
\bea
W &=& S + [p_\tau \tau] |^{t''}_{t'} \\
&=& \int [p\d_t q + \tau \d_t p^\tau - N(H-p^\tau)] dt. \nonumber
\eea
Integrating the equation of motion $ \d_t p^\tau =0$
gives $p^\tau = E$ and leads to the reduced action
\beq
\label{action_charac}
W = S + E(\tau''-\tau') = \int_{t'}^{t''} [p\d_t q  - N(H-E)] dt.
\eeq
This is Hamilton's characteristic function \cite{Gol}, and
it is equivalent to Jacobi's action \cite{Lan}.

This form of the action (\ref{action_charac}) is similar to 
(\ref{action_param}) but differs significantly by a total derivative 
(surface term). $N$ is now free to be varied, and the correct result 
$H=E$ is found.  This is the ``integrated'' form of Noether's theorem for time 
reparametrization invariance.

To continue this analysis in field theory, we begin in 
a basis adapted to the foliation and then parametrize 
the metric.  For an ADM-like treatment, the metric is (partially) 
parametrized as 
\bea
\label{metric}
ds^2 &=& -(\w^0)^2 + g_{ab} \w^a \w^b \\
&=& -N^2 (dt)^2 + g_{ij} (dx^i + \beta^i dt)(dx^j + \beta^j dt), \nonumber
\eea
where $N$ is the lapse and $\beta^i$ is the spatial shift vector.  
Clearly, parametrizing the metric is parametrizing the basis 1-forms, 
\beq
\label{param_0}
\om^{\al}= n_\mu^{(\al)} dx^\mu,
\eeq
where ${\al} = (0,a)$, $\mu = (t,i)$, and $dx^\mu$ is the basis 1-form of a 
coordinate frame but could be a general 1-form.  This is the analog
of $d\tau = N(t) dt$ in the particle case.  It is more general than
\Kuchar's embedding variable description because $\om^{\al}$ may
not be the exterior derivative of a scalar coordinate function.  Note
that $N$ and $\beta^k$ simply adjust the coordinate description of the 
reference frame $\om^{\al}$, thus the frame is fixed in the sense that
it does not change when $N$ and $\beta^k$ are varied. 

The coordinate-free condition one must impose is 
\beq
\label{param}
e_{\al}( \om^{\al}-n_\mu^{(\al)} dx^\mu) \equiv 
d\cP - e_{\al} n_\mu^{(\al)} dx^\mu
=0.
\eeq
Here, 
\beq
d\cP \equiv e_{\al} \om^{\al}
\eeq
is Cartan's unit tensor\cite{Car}, where $\cP$ may be understood as a point
on the manifold and $d\cP$ is loosely its exterior derivative.
While not strictly speaking an exterior derivative, $d\cP$ shares 
certain properties with true exterior derivatives.  In particular,
$d^2 \cP$ vanishes (equivalent to vanishing torsion)
while under the action of an infinitesimal diffeomorphism
\beq
\label{inf_diffeo}
\delta_\xi d\cP \equiv \Lie_\xi (e_a \w^a) 
=d (e_a \xi^a) \equiv d \delta_\xi \cP.
\eeq
where $\Lie_\xi= d i_\xi + i_\xi d$ is the Lie derivative along
the spacetime vector $\xi$ and $i_\xi$ is the interior product with $\xi$.
Thus, an active infinitesimal diffeomorphism acts by moving points $\cP$ on 
the manifold.

The parametrization condition is imposed using a Lagrange multiplier
$\star \bT *$, the double dual of the $(1,1)$ form 
$\bT =e_\al T^\al\mathstrut_\be \w^\be$,
where $*$ is the Hodge dual for forms and $\star$ is its analog for
antisymmetric products of vectors.  In components, one has
$$\star \bT * = (1/36) e_\gam \wedge e_\delta \wedge e_\lambda 
\eps^{\al \gam \delta \lambda} T_\al\mathstrut^\be
\eps_{\be \rho \sig \tau} \w^\rho \wedge \w^\sig \wedge \w^\tau.$$
The parametrization-fixing action is 
\beq
\label{action_p}
S_p = \int \star[{\star \bT *} \wedge (d\cP - e_{\al} n^{(\al)}\mathstrut_\mu
dx^\mu)].
\eeq
(The second overall $\star$ converts an antisymmetric product of four basis
vectors to a scalar.)  In coordinates this is
\beq
S_p = \int T_\al\mathstrut^\be d^3\Sig_\be \wedge 
(\om^\al - n^{(\al)}_\mu dx^\mu),
\eeq
where $d^3\Sig_\be = (1/6) \eps_{\be \rho \sig \tau} \w^\rho \wedge \w^\sig 
\wedge \w^\tau$.  

Before imposing parametrization with (\ref{action_p}), simply 
change variables between the different forms of the metric (\ref{metric}) in
the  covariant field theory action,
\beq
\label{Lagrangian}
S_M= \int \cL \,\om^4= \int [\pi^A \d_{0} \phi_A - \cH] \om^4,
\eeq
where $\cL$ is the Lagrangian density of the field, 
$A$ is a multi-index allowing for a general set of fields
with arbitrary index structure, 
$\pi^A\equiv \delta \cL /\delta \d_{0} \phi_A$ is the
momentum conjugate to $\phi_A$, and
$\om^4 = \om^{0}\wedge \om^{1} \wedge \om^{2} \wedge
\om^{3}$. Note that $\cL$ is a four-density
while $\pi$ and $\cH$ are three-densities. In the foliation-adapted metric,
the action becomes
\beq
\label{action_t}
S_M = \int [\pi^A \d_t \phi_A - N \cH - \beta^k \cH_k] d^4 x
\eeq
where $ N \d_{0}=\d_t - \beta^k \d_k$ ($\d_k= \d/\d x^k$), and
$\cH$ and $\cH_k = \pi^A \d_k \phi_A$ are three-densities.

Were $N$ varied in (\ref{action_t}), we would
find the absurd result that the energy density vanishes, $\cH=0$; this is an
error \cite{Kuc81}.  Similarly, variation of $\beta^k$ in (\ref{action_t})
would imply that the momentum density vanishes $\cH_k=0$. 
As in the particle example, the problem is that $N$ and $\beta^k$ are
not truly free.  They arise in parametrizing the metric through
(\ref{param}), and proper account of this fact must be taken. 
\Kuchar\ attempts this through his introduction of the embedding 
variables\cite{Kuc76,Kuc81}, but this puts an emphasis on coordinate 
functions that can lead one to misjudge the number of
parameters introduced in (\ref{param_0}).

Varying $N$ and $\beta^k$ is a passive spacetime diffeomorphism since they
merely change coordinates on a fixed background.  If instead an 
infinitesimal variation (active spacetime diffeomorphism) were made directly 
on the metric dependence in (\ref{Lagrangian}),
\beq
\label{inf_diffeo2}
\delta_\xi g^{\al\be} = - \grad^{\al} \xi^{\be} - \grad^\be \xi^\al,
\eeq
the correct result of Noether's theorem would be found
\beq
\label{Noether}
\grad^\al {\delta \cL\over \delta g^{\al \be}} =0.
\eeq
As in the particle analog---$dH/dt=0$---an integration ``constant,''
or zero-mode, may be added to this.  Here it is a symmetric 
divergence-free tensor density, the stress-tensor $T_{\al \be}$,
\beq
\label{Noether2}
\grad^\al ({\delta \cL\over \delta g^{\al \be}} 
+ {1\over 2} (-\ddet4g)^{1/2} T_{\al \be})
=0,
\eeq
where $\mathstrut^4 g = \det \mathstrut^4 g_{\al \be}$ is the
determinant of the 4-metric.
(Of course, $T_{ab}$ generally isn't constant.)  This result
may be obtained directly by varying
the parametrization-fixed action $ S = S_M - S_p$
using (\ref{inf_diffeo}) and (\ref{inf_diffeo2}).

Fixing the parametrization with (\ref{action_p}) makes the 
Hamiltonian action
\beqa
\label{action_fixed}
S= S_M - S_p 
&=& \int  [\pi^A \d_t \phi_A - N \Ham - \beta^k \Ham_k] d^4 x  \\
&& \hspace{0.5cm} -\int \star[{\star \bT *} \wedge 
(d\cP - e_{\al} n^{(\al)}\mathstrut_\mu dx^\mu)]. \nonumber
\eeqa
Use the parametrization implied by (\ref{metric}):  $n^{(0)}\mathstrut_t = N$,
$n^{(a)}\mathstrut_j=\delta^a\mathstrut_j$, and
$n^{(a)}\mathstrut_t = \delta^a\mathstrut_k \beta^k$,
with all others zero. Variation of $N$ leads to
\beq
\label{energy}
\Ham - T^0\mathstrut_0 g^{1/2}= 0
\eeq
while variation of $\beta^k$ leads to
\beq
\label{mom}
\Ham_k - T^0\mathstrut_k N g^{1/2} =0.
\eeq
The energy and momentum densities no longer vanish but equal the appropriate
stress tensors.  Since the metric is only partially parametrized, only 
the energy and momentum densities arise.   

Covariance of the Lagrangian guarantees that the other components of 
the stress tensor will be found by parametrizing the other components of 
the metric.  The non-covariant form of the Hamiltonian action hides this 
fact, but it is easily verified by using a covariant canonical
formalism \cite{Kuc76,Schw} and suitably parametrizing a background 
orthonormal frame.

The situation in gravity can now be explained.  The traditional ADM
Hamiltonian formulation of general relativity uses the 
partially parametrized metric (\ref{metric}) and obtains an action
of the form (\ref{action_t}) with $\phi_A$ replaced by $g_{ij}$ and
$\pi^A$ by the momentum conjugate to the metric, $\pi^{ij}$.  For this
reason, the action for relativity is sometimes referred to as
``already parametrized.''  When
the lapse and shift are varied, the Hamiltonian (energy) and momentum densities
are found to vanish.  The lapse and shift are not free, however, but
restricted by their role in the parametrization.   
This parametrization must be fixed using a Lagrange multiplier as 
in (\ref{action_p}).  The results are the same as in field theory
(\ref{Noether2})-(\ref{mom}). The action is not ``already'' parametrized
({\it cf.} also \cite{Tor}).  The Hamiltonian and momentum
densities are specified initially and then 
propagated by the Bianchi identities (\ref{bi1}) \cite{AnY98}.

Furthermore, because the original Hilbert action is covariant, the entire
metric should be parametrized.  It is not enough to parametrize only
the ``time'' direction using the lapse and shift.  This implies 
that, from the standpoint of the passive transformations of the 
parametrized theory, the entire metric simply fixes a reference background 
(just as it does in field theory).  This point is made forcefully in a tetrad 
formulation of gravity, where the background orthonormal metric is 
clearly non-dynamical.

Finally, the term $\int \star[\star \bT *\wedge d\cP]$ in (\ref{action_fixed})
looks like it is 
a surface integral because $d(\star \bT *)=0$, but unfortunately
$d\cP$ is only formally an exterior derivative.  Thus, while
one is tempted to use Routh's procedure to reach a Jacobi
form of the action, this cannot be done.  A reason underlying
this is that $\grad^\mu T_{\mu\nu}=0$ cannot be integrated to
a conservation law for energy and momenta unless $T_{\mu\nu}$ is 
contracted with an appropriate Killing vector $\xi^\nu$.  Such a Killing 
vector is not always available, and gravity is generically
time-dependent. 

I thank J.W. York, Jr. for his consistent support and encouragement. 
This work was supported in part by National Science Foundation grant 
PHY-9413207.

\end{document}